\documentclass[twocolumn,showpacs]{revtex4}
\usepackage{graphicx}
\usepackage{dcolumn}
\usepackage{bm}
\usepackage{amsmath}
\usepackage{amsfonts}
\usepackage{amssymb}
\providecommand{\U}[1]{\protect\rule{.1in}{.1in}}

\begin{document}
\title{Distribution of chirality in the quantum walk: Markov process and entanglement}
\author{Alejandro Romanelli}
\altaffiliation{alejo@fing.edu.uy}
\affiliation{Instituto de F\'{\i}sica, Facultad de Ingenier\'{\i}a\\
Universidad de la Rep\'ublica\\
C.C. 30, C.P. 11000, Montevideo, Uruguay}
\date{\today }

\begin{abstract}
The asymptotic behavior of the quantum walk on the line is
investigated focusing on the probability distribution of chirality
independently of position. It is shown analytically that this
distribution has a long-time limit that is stationary and depends on
the initial conditions. This result is unexpected in the context of
the unitary evolution of the quantum walk, as it is usually linked
to a Markovian process. The asymptotic value of the entanglement
between the coin and the position is determined by the chirality
distribution. For given asymptotic values of both the entanglement
and the chirality distribution it is possible to find the
corresponding initial conditions within a particular class of
spatially extended Gaussian distributions.

\end{abstract}

\pacs{03.67.-a, 03.65.Ud, 02.50.Ga}
\maketitle

\section{Introduction}
The quantum walk (QW) on the line \cite{QW} is a natural
generalization of the classical random walk in the frame of quantum
computation and quantum information processing and it is receiving
permanent attention\cite{childs,Linden,Alejo3}. It has the property
to spread over the line linearly in time as characterized by the
standard deviation $\sigma (t)\sim t$, while its classical analog
spreads out as $\sigma (t)\sim t^{1/2}$. This property, as well as
quantum parallelism and quantum entanglement, could be used to
increase the efficiency of quantum algorithms. As an example the QW
has been used as the basis for optimal quantum search algorithms
\cite{Shenvi,Childs et} on several topologies. On the other hand,
some experimental implementations of the QW have been reported
\cite{Exp}, and others have been proposed by a number of authors
\cite{ProExp}.

The concept of entanglement is an important element in the
development of quantum communication, quantum cryptography and
quantum computation. In this context, several authors have studied
the QW subjected to different types of coin operators and/or sources
of decoherence to analyze the long-time entanglement between the
coin and the position and its relation with the initial conditions.
Carneiro \emph{et al.} \cite{Carneiro} investigate entanglement
between the coin and the position calculating the entropy of the
reduced density matrix of the coin. The relation between asymptotic
entanglement and nonlocal initial conditions (in the one and two
dimensional QW) is treated in
\cite{abal,Annabestani,Omar,Pathak,Petulante,Venegas}. Ref.
\cite{Endrejat,Ellinas1,Ellinas2} analyze the effect of entanglement
on the initial coin state, which is measured by the mean value of
the walk, and the relation between the entanglement and the symmetry
of the probability distribution. In Ref. \cite{Maloyer} the relation
between entanglement and decoherence is studied numerically. Refs.
\cite{Venegas1,Chandrashekar} propose to use the QW as a tool for
quantum algorithm development and as entanglement generator,
potentially useful to test quantum hardware.

In the previous works
\cite{Carneiro,abal,Annabestani,Omar,Pathak,Petulante,Venegas,Endrejat,Ellinas1,Ellinas2,Maloyer,Venegas1,Chandrashekar}
the QW evolution was studied using the amplitude of probability to
evaluate the dynamics. In this work we introduce a new probability
distribution, the global chirality distribution (GCD) that is the
distribution of chirality independently of the walker's position. We
show that the GCD has an asymptotic limit and we connect this limit
with the entropy of entanglement between the coin and the position.
The asymptotic behavior of GCD is an unexpected result because, due
to the unitary evolution, the QW does not converge to any stationary
state as would be the case \emph{e.g.} of a Markov chain.  In order
to show these results, we rewrite the QW evolution equation as the
sum of two different terms, one responsible for the classical
diffusion and the other for the quantum coherence \cite{Alejo1}. As
we shall see, the first term above obeys a master equation as is
typical of Markovian processes, while the second term includes the
interference needed to preserve the unitary character of the quantum
evolution. This approach provides a more intuitive framework which
proves useful for analyzing the behavior of quantum systems with
decoherence. It allows to study the quantum evolution together with
the associated classical Markovian process at all times, and in
particular the asymptotic behavior of the GCD.

The paper is organized as follows. In the next section we develop
the standard QW model, in the third section we build the master
equation for the GCD, in the fourth section we present the
asymptotic solution for the QW, in the fifth section the entropy of
entanglement is connected with the GCD, and in the last section we
draw the conclusions.

\section{The standard QW}

The QW on the line, corresponds to a one-dimensional evolution of a quantum
system in a direction which depends on an additional degree of freedom, the
chirality, with two possible states: \textquotedblleft
left\textquotedblright\ $|L\rangle $\ or \textquotedblleft
right\textquotedblright\ $|R\rangle $. The global Hilbert space of the
system is the tensor product $H_{s}\otimes H_{c}$ where $H_{s}$ is the
Hilbert space associated to the motion on the line and $H_{c}$ is the
chirality Hilbert space. Let us call $T_{-}$ ($T_{+}$) the operators in $%
H_{s}$ that move the walker one site to the left (right), and $|L\rangle
\langle L|$ and $|R\rangle \langle R|$ the chirality projector operators in $%
H_{c}$. We consider the unitary transformations
\begin{equation}
U(\theta )=\left\{ T_{-}\otimes |L\rangle \langle L|+T_{+}\otimes |R\rangle
\langle R|\right\} \circ \left\{ I\otimes K(\theta )\right\} ,  \label{Ugen}
\end{equation}%
where $K(\theta )=\sigma _{z}\cos\theta+i\sigma _{x}\sin\theta$, $I$ is the
identity operator in $H_{s}$, and $\sigma _{z}$ and $\sigma _{x}$ are Pauli
matrices acting in $H_{c}$. The unitary operator $U(\theta )$ evolves the
state in one time step as
\begin{equation}
 |\Psi (t+1 )\rangle =U(\theta )|\Psi (t)\rangle .  \label{evolution}
\end{equation}
The wave vector can be expressed as the spinor
\begin{equation}
|\Psi (t)\rangle =\sum\limits_{k=-\infty }^{\infty }\left[
\begin{array}{c}
a_{k}(t) \\
b_{k}(t)%
\end{array}%
\right] |k\rangle ,  \label{spinor}
\end{equation}%
where the upper (lower) component is associated to the left (right)
chirality. Substituting Eq. (\ref{spinor}) and Eq. (\ref{Ugen}) into
Eq. (\ref{evolution}) and projecting over the position vector
$|k\rangle $ the unitary evolution is written as the map
\begin{align}
a_{k}(t+1 )& =a_{k+1}(t)\,\cos \theta \,+b_{k+1}(t)\,\sin \theta , \,  \notag
\\
b_{k}(t+1 )& =a_{k-1}(t)\,\sin \theta \,-b_{k-1}(t)\,\cos \theta .
\label{mapa}
\end{align}

\section{unitary evolution and master equation for the chirality}

In references \cite{Alejo1, Alejo2}, it is shown how a unitary
quantum mechanical evolution can be separated into Markovian and
interference terms. Here we use this method to recognize a master
equation in chirality starting from the original map Eq.
(\ref{mapa}). First we define the left and right distributions of position as $%
P_{kL}(t)=\left| a_{k}(t)\right| ^{2}$ and $P_{kR}(t)~\equiv~\left|
b_{k}(t)\right|^{2}$ respectively. Combining the two components of
the Eq. (\ref{mapa}) and after some simple algebra we obtain
\begin{align}
P_{k,L}(t+1) & =P_{k+1,L}(t)\,\cos^{2}\theta+P_{k+1,R}(t)\,\sin^{2}\theta
\notag \\
& +\beta_{k+1}(t) \,\sin2\theta ,  \notag \\
P_{k,R}(t+1) & =P_{k-1,L}(t)\,\sin^{2}\theta+P_{k-1,R}(t)\,\cos^{2}\theta
\notag \\
& -\beta_{k-1}(t) \,\sin2\theta\,,  \label{mapa1}
\end{align}
where $\beta _{k}\equiv\mathrm{Re}\left[
a_{k}(t)b_{k}^{\ast}(t)\right] $ is an interference term, with
$\mathrm{Re}(z)$ indicating the real part of $z$. Of course the
probability distribution for the position is
$P_{k}(t)=P_{kL}(t)+P_{kR}(t)$. We define the global left and right
chirality probabilities as
\begin{align}
P_{L}(t)\equiv \sum_{k=-\infty }^{\infty } P_{kL}(t)=\sum_{k=-\infty
}^{\infty }\left\vert a_{k}(t)\right\vert ^{2} , \,  \notag \\
P_{R}(t)\equiv \sum_{k=-\infty }^{\infty } P_{kR}(t)=\sum_{k=-\infty
}^{\infty }\left\vert b_{k}(t)\right\vert ^{2} ,  \label{chirality}
\end{align}
with $P_{R}(t)+P_{L}(t)=1$ and the global chirality distribution
(GCD) is defined as the distribution formed by the couple
$\left[
\begin{array}{c}
P_{L}(t) \\
P_{R}(t)%
\end{array}%
\right]$.

Using the definition Eq. (\ref{chirality}) in Eq. (\ref{mapa1}) we
have
\begin{align}
{\left[
\begin{array}{c}
P_{L}(t+1) \\
P_{R}(t+1)%
\end{array}%
\right]} & = {\left(
\begin{array}{cc}
\cos^{2}\theta & \sin^{2}\theta \\
\sin^{2}\theta & \cos^{2}\theta%
\end{array}%
\right)} \left[
\begin{array}{c}
P_{L}(t) \\
P_{R}(t)%
\end{array}%
\right]  \notag \\
& +\mathrm{Re}\left[Q(t)\right]\sin{2}\theta\left[
\begin{array}{c}
1 \\
-1%
\end{array}%
\right],  \label{master}
\end{align}
where
\begin{equation}
Q(t)\equiv \sum_{k=-\infty }^{\infty } a_{k}(t)b_{k}^{\ast}(t).
\label{qdet}
\end{equation}
In Eq. (\ref{master}) the two dimensional matrix can be interpreted
as a transition probability matrix for a classical two dimensional
random walk as it satisfies the necessary requirements, namely, all
its elements are positive and the sum over the elements of any
column or row is equal to one. On the other hand, it is clear that
$Q(t)$ accounts for the interferences. When $Q(t)$ vanishes the
behavior of the GCD can be described as a classical Markovian
process. However $Q(t)=0$ does not necessarily imply the loss of
unitary evolution; such a loss requires the vanishing of all the
$\beta_{k}(t)$ (see Eq (\ref{mapa1})). As shown in Ref. \cite
{Alejo4} the primary effect of decoherence is to make the
interference terms $\beta_{k}(t)$ negligible; in this case Eq.
(\ref{mapa1}) becomes a true master equation. On the other hand,
when $Q(t)$ is time independent, that is $Q(t)=Q=constant$, then Eq.
(\ref{master}) is solved using the methods developed in \cite{Cox};
its solution as a function of the initial GCD is
\begin{eqnarray}
{\left[
\begin{array}{c}
P_{L}(t) \\
P_{R}(t)%
\end{array}%
\right] } &=&\frac{1}{2}{\left(
\begin{array}{cc}
1+\cos ^{t}2\theta  & 1-\cos ^{t}2\theta  \\
1-\cos ^{t}2\theta  & 1+\cos ^{t}2\theta
\end{array}%
\right) }\left[
\begin{array}{c}
P_{L}(0) \\
P_{R}(0)%
\end{array}%
\right]   \notag \\
&&+\mathrm{Re}\left[ Q \right] \frac{1-\cos ^{t}2\theta }{\tan \theta }%
\left[
\begin{array}{c}
1 \\
-1%
\end{array}%
\right] .  \label{markovian}
\end{eqnarray}%
Taking the limit $t\rightarrow \infty $ in Eq. (\ref{markovian}) it
is possible to obtain the asymptotic value of the GCD as a function
of its initial value and $Q$.

In the generic case $Q(t)$ is a time depend function but in this
system (as will be seen in the next section) $Q(t)$, $P_{L}(t)$ and
$P_{R}(t)$ have long-time limiting values which are determined by
the initial conditions of Eq. (\ref{mapa}). Therefore we can solve
Eq. (\ref{master}) in this limit, defining
\begin{align}
\Pi _{L}\equiv P_{L}(t\rightarrow \infty ), \,  \notag
\\
\Pi _{R}\equiv P_{R}(t\rightarrow \infty ), \,  \notag
\\
Q_{0}\equiv Q(t\rightarrow \infty ), \,  \label{asym}
\end{align}
and substituting these asymptotic values in Eq. (\ref{master}), to
obtain the stationary solution for the GCD
\begin{equation}
{\left[
\begin{array}{c}
\Pi _{L} \\
\Pi _{R}%
\end{array}%
\right] }=\frac{1}{2}\left[
\begin{array}{c}
1+2\mathrm{Re}(Q_{0})/\tan \theta  \\
1-2\mathrm{Re}(Q_{0})/\tan \theta
\end{array}%
\right]. \label{estacio}
\end{equation}%
This interesting result for the QW shows that the long-time
probability to find the system with left or right chirality has a
limit.
%As we shall see below, this limit depends on the initial
%values of the amplitudes $a_{k}$ and $b_{k}$ through $Q_{0}$.

In the next section we show that it is possible to have $Q_{0}=0$
choosing adequately the initial conditions. In this case,
Eq.(\ref{master}) approaches a Markov chain \cite {Cox} with two
states and the dynamics of the GDC turns into an example of
dependent Bernoulli trials in which the probabilities of success or
failure et each trial depend on the outcome of the previous trial.
Now the only asymptotic solution is $\Pi _{L}=\Pi _{R}=1/2$ (see
Eq.(\ref{estacio})).

If we look back at Eq.(\ref{evolution}) in connection with
Eq.(\ref{estacio}) a paradoxical situation arises. The dynamical
evolution of the QW is unitary but the evolution of its GCD has an
asymptotic value characteristic of a diffusive behavior. This
situation is further surprising if we compare our case with the case
of the QW on finite graphs \cite{Aharonov} where it is shown that
there is no converge to any stationary distribution.

\section{Asymptotic solution for the QW}

In previous works \cite{Alejo2, Alejo3} an alternative analytical approach
was presented to obtain the asymptotic behavior of the QW on the line. The
discrete map was substituted by two continuous differential equations for $%
a_{k}(t)$ and $b_{k}(t)$ starting from a characteristic time
$t_0>>1$. The initial conditions for these equations are not
necessarily the same as those used in the discrete map Eqs.
(\ref{mapa}), because the approximation of a finite difference by a
derivative does not hold for small times. However these initial
conditions must assure the same asymptotic behavior than that of the
discrete map.

The asymptotic solutions of Eqs. (\ref{mapa}) given by the
differential equations are
\begin{align}
a_{k}(t)\simeq \sum\limits_{l=-\infty }^{\infty }\left( -1\right)
^{k-l}a_{l}^{0}\text{ }\,J_{k-l}(t\, \cos \theta) , \,  \notag \\
b_{k}(t)\simeq \sum\limits_{l=-\infty }^{\infty }\left( -1\right)
^{k-l}b_{l}^{0}\text{ }\,J_{k-l}(t\, \cos \theta),  \label{solub}
\end{align}
where $J_{l}$ is the $l$th order cylindrical Bessel function and $a_{k}^{0}$
and $b_{k}^{0}$ are initial amplitudes for the differential equations. To
secure that the behavior of the discrete map and the differential equations
are the same in the asymptotic regime we should choose ${a_{k}^{0}}$ and ${%
b_{k}^{0}}$ to be smoothly extended in space.

Replacing Eq. (\ref{solub}) in Eqs. (\ref{chirality},\,\ref{qdet})
and noting that the Bessel functions satisfy  $\sum_{j=-\infty
}^{\infty }J_{j}(t)J_{j-k }(t)=\delta _{k 0}$ we have
\begin{equation}
Q(t)=\sum_{k=-\infty }^{\infty }a_{k}^{0}b_{k}^{0\ast }=Q_0 ,
\label{a0}
\end{equation}
\begin{equation}
P_{L}(t)=\sum_{j=-\infty }^{\infty }\left\vert a_{k}^{0}\right\vert
^{2}=\Pi_L, \label{a}
\end{equation}
\begin{equation}
P_{R}(t)=\sum_{k=-\infty }^{\infty }\left\vert b_{k}^{0}\right\vert
^{2}=\Pi_R. \label{c}
\end{equation}
The time independence of $Q(t)$, $P_{L}(t)$ and $P_{R}(t)$ is a
consequence of the asymptotic approach given by Eq. (\ref{solub})
and evidently their
values are $Q_0$, $\Pi_{L}$ and $\Pi_{R}$ respectively. When $Q(t)$, $%
P_{L}(t)$ and $P_{R}(t)$ are calculated with the map Eq.
(\ref{mapa}), they have a transient time dependence (for $t<t_0$)
after which they attain their asymptotic values $Q_0$, $\Pi_L$ and
$\Pi_R$, as shown in \cite{Alejo2}.

We propose for the initial conditions $a_{k}^{0}$ and $b_{k}^{0}$ the
following extended Gaussian distributions \cite{Eugenio}
\begin{equation}
a_{k}^{0}\equiv\left\{ \frac{1}{\sigma_0\sqrt{2\pi }}\exp \left[ -\frac{%
(k-k_0)^{2}}{2\sigma_0^{2}}\right] \right\} ^{\frac{1}{2}}\cos \alpha \text{,%
}  \label{aes}
\end{equation}%
\begin{equation}
b_{k}^{0}\equiv a_{k}^{0}\, \tan\alpha \,\exp (i \delta ) \text{,}
\label{bes}
\end{equation}%
where $\sigma_0$ is the initial standard deviation, $k_0$ is the central
position of the Gaussian distribution, $\alpha$ is a parameter that
determines the initial proportion of the left and right chiralities and $%
\delta$ is a phase to be determined bellow as a function of $\alpha$
and $\theta$.
\begin{figure}[th]
\begin{center}
\includegraphics[scale=0.38]{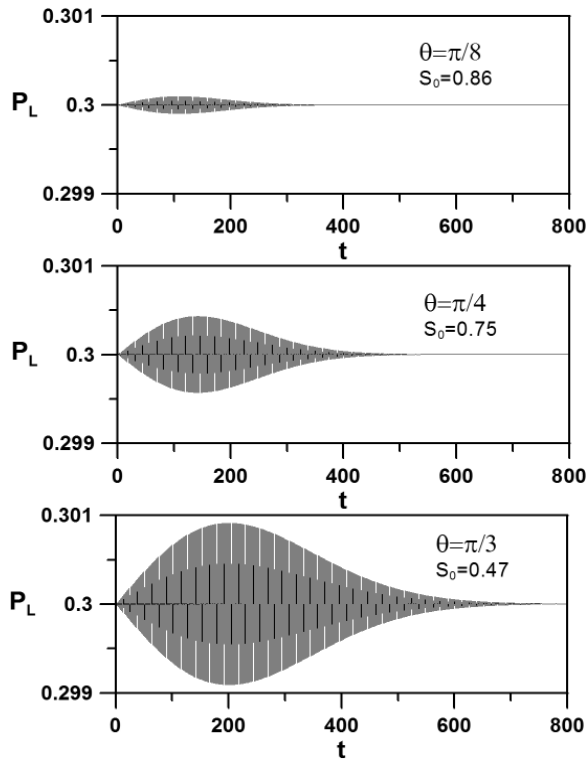}
\end{center}
\caption{The global left probability $P_L(t)$ as a function of the dimensionless time $t$
calculated with the map Eq. (\protect\ref%
{mapa}). The initial conditions are given by Eqs.
(\protect\ref{aes},\,\protect
\ref{bes}). It is shown for three values of $\protect\theta$, ${\protect\sigma%
_0=100}$, $k_0=0$ and $\cos^2\protect\alpha=0.3$. The approximate values of $%
t_0$ are, from top to bottom, $300,500,700$. The asymptotic values
for the dimensionless entropy $S_0$ are also presented. } \label{f1}
\end{figure}
\begin{figure}[th]
\begin{center}
\includegraphics[scale=0.38]{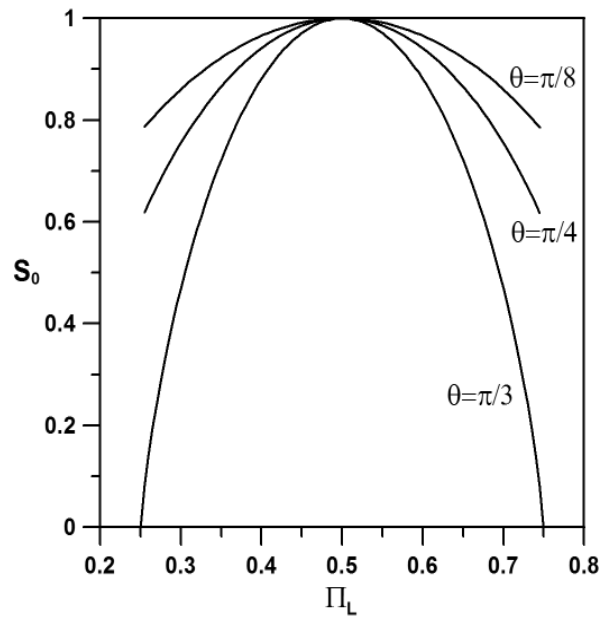}
\end{center}
\caption{The dimensionless entropy of entanglement $S_0$ as a
function of the asymptotic global left probability $\Pi_L$ for three
different values of $\protect\theta$.} \label{f2}
\end{figure}

Now we evaluate $Q_0$, $\Pi_L$ and $\Pi_R$ using the initial
conditions Eqs. (\ref{aes},\,\ref{bes}) in Eqs.
(\ref{a0},\,\ref{a},\,\ref{c}) and noting that
 $\sum_{k=-\infty }^{\infty }\exp \left[ -\frac{%
k^{2}}{2\sigma_0^{2}}\right]\cong \sqrt{2\pi} \sigma_0$, for
$\sigma_0\gg1$
\begin{equation}
Q_0=\frac{1}{2}\sin2\alpha\,\cos\delta\,,  \label{q0}
\end{equation}%
\begin{equation}
\Pi_L=\cos^2\alpha\,,  \label{pl}
\end{equation}
\begin{equation}
\Pi_R=\sin^2\alpha\,.  \label{pr}
\end{equation}%
On the other hand, from Eq. (\ref{estacio}) we see that $Q_0$ and
$\Pi_L$ are not independent, then substituting Eqs.
(\ref{q0},\,\ref{pl},\,\ref{pr}) into Eq. (\ref{estacio}) we have
\begin{equation}
\cos\delta=\frac{\tan\theta}{\tan2\alpha}\,.  \label{beta}
\end{equation}
Then, we rewrite Eq. (\ref{q0}) as a function of the two independent
parameters of the model $\theta$ and $\alpha$
\begin{equation}
Q_0=\frac{1}{2}\cos2\alpha\,\tan\theta\,,  \label{q1}
\end{equation}
note that $Q_0$ vanishes for $\alpha=\pi/4$.
\begin{figure}[th]
\begin{center}
\includegraphics[scale=0.38]{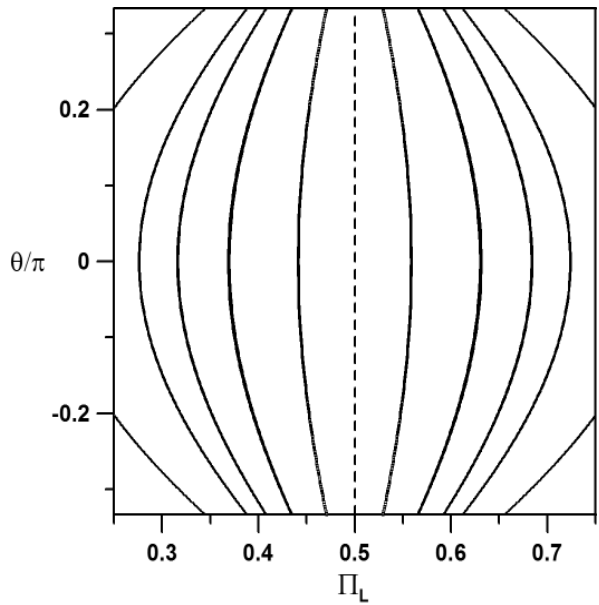}
\end{center}
\caption{Level curves for the dimensionless entropy of entanglement
$S_0$ as a function of the dimensionless angle
$\frac{\protect\theta}{\pi}$ and the asymptotic global left
probability $\Pi_L$. Five curves, in full line, are presented; each
curve has two branches placed symmetrically on both sides of the
central straight dashed line $\Pi_L=0.5$. Starting from this line,
where $S_0=1$, the values of $S_0$ are $0.99, 0.95, 0.90, 0.85$ and
$0.70$.} \label{f3}
\end{figure}
In order to verify the approximations made in our analytical
treatment we shall compare the result of Eq. (\ref{pl}) with the
numerical evaluation of the asymptotic behavior of $P_L(t)$ using
the map Eq. (\ref{mapa}) and the initial conditions given by Eqs.
(\ref{aes},\,\ref{bes}). This calculation is presented in
Fig.~\ref{f1}. Our selection of the initial amplitudes makes that
the asymptotic value for $P_L(t)$ is equal to the initial one. Which
in turn is the same as that given by Eq. (\ref{pl}). Thus the
asymptotic behaviors of Eq. (\ref{mapa}) is in excellent agreement
with our theoretical approach. Our treatment works very well for
values of $\sigma_0> 10$. The asymptotic regime of $P_L(t)$ sets in
at the time $t_0$ after some strong oscillations. The value of $t_0$
depends on the parameters of the problem.

To sum up, the dynamical evolution of the QW is defined by Eq.
(\ref{mapa}), but it is possible to obtain a predetermined
asymptotic value of the GCD $\left[\begin{array}{c}\Pi_{L} \\
\Pi_{R}\end{array}\right]$ using as initial conditions Eqs.
(\ref{aes},\ref{bes}), where the parameters $\alpha$ and $\delta$
are determined by the Eqs. (\ref{pl},\ref{beta}) and the parameters
$k_0$ and $\sigma_0$ are free to be adjusted.

\section{Entropy of entanglement}
The unitary evolution of the QW generates entanglement between the
coin and
position degrees of freedom. This entanglement will be characterized \cite%
{Carneiro,abal} by the von Neumann entropy of the reduced density operator,
called entropy of entanglement. The quantum analogue of the Shannon entropy
is the von Neumann entropy
\begin{equation}
S_{N}(\rho )=-\mathrm{tr}(\rho \log \rho ).  \label{uno}
\end{equation}%
where $\rho =|\Psi (t)\rangle \langle \Psi (t)|$ is the density matrix of
the quantum system. Due to the unitary dynamics of the QW the system remains
in a pure state and this entropy vanishes. However, for these pure states,
the entanglement between the chirality and the position can be quantified by
the associated von Neumann entropy for the reduced density operator
\begin{equation}
S(\rho )=-\mathrm{tr}(\rho _{c}\log \rho _{c}).  \label{dos}
\end{equation}
where $\rho _{c}=\mathrm{tr}(\rho )$ and the partial trace is taken
over the positions. Using the wave function Eq. (\ref{spinor}) and
its normalization properties, the reduced density operator is
explicitly expressed as
\begin{equation}
\rho_{c} =\left(
\begin{array}{cc}
P_{L}(t) & Q(t) \\
Q(t)^{\ast } & P_{R}(t)%
\end{array}%
\right) .  \label{rho}
\end{equation}
The reduced entropy can be expressed through the two eigenvalues $\left\{
\lambda _{+},\,\lambda _{-}\right\} $ of the reduced density matrix as
\begin{equation}
S(t)=-\lambda _{+}\log _{2}\lambda _{+}-\lambda _{-}\log _{2}\lambda _{-}.
\label{ttres}
\end{equation}%
The expressions for the eigenvalues are
\begin{equation}
\lambda _{\pm}=\frac{1}{2}\left[ 1\pm \sqrt{1+4\left( \left\vert
Q(t)\right\vert ^{2}-P_L(t)\,P_R(t)\right) }\right].  \label{lam}
\end{equation}
In the asymptotic regime $\lambda _{\pm}\rightarrow\Lambda _{\pm}$ where
\begin{equation}
\Lambda _{\pm}=\frac{1}{2}\left[ 1\pm \sqrt{1+4\left( \left\vert
Q_0\right\vert ^{2}-\Pi_L\,\Pi_R\right) }\right],  \label{lam1}
\end{equation}
and the corresponding entropy ($S(t)\rightarrow S_0$) is
\begin{equation}
S_0=-\Lambda _{+}\log _{2}\Lambda _{+}-\Lambda _{-}\log _{2}\Lambda _{-}.
\label{s0}
\end{equation}
Using the initial conditions Eqs. (\ref{aes},\,\ref{bes}) in Eq.
(\ref{lam1}) we have
\begin{equation}
\Lambda _{\pm}=\frac{1}{2}\left[ 1\pm \frac{\cos2\alpha}{\cos\theta}\right].
\label{lam2}
\end{equation}

For $\alpha=\pi/4$ both eigenvalues are $\Lambda _{\pm}=1/2$ and from Eqs. (%
\ref{pl},\,\ref{pr},\,\ref{q1}) $\Pi_L=\Pi_R=1/2$ and $Q_0=0$. For
this value the entropy of entanglement Eq. (\ref{s0}) has its
maximum value $S_0=1$. Therefore the maximum value of the entropy of
entanglement is achieved for the classical Markovian process
($Q_0=0$). Note that this result is true for all initial conditions
that satisfy $Q(t)\rightarrow 0$, as it follows from Eqs.
(\ref{master},\,\ref{estacio},\,\ref{lam1},\,\ref{s0}).

For $\alpha=\theta/2$, the entropy attains its minimum value
$S_0=0$, see Eqs. (\ref{s0},\,\ref{lam2}). Then in this case there
is no entanglement between coin and position.

Using the results of the previous section it is clear that starting from
given initial conditions the asymptotic values $\Pi_L$ and $Q_0$ are
obtained, and then the entropy of entanglement is calculated using Eqs. (\ref%
{lam1},\,\ref{s0}). The inverse path is also possible, that is
starting from a predetermined value of the entropy of entanglement
Eq. (\ref{s0}) it is possible to obtain the initial conditions Eqs.
(\ref{aes},\,\ref{bes}) of the system that produce this entanglement
asymptotically.

The previous ideas are numerically implemented using Eqs.
(\ref{estacio},\,\ref{lam1},\,\ref{s0}) and taking $Q_0$ as a real
constant, and the results are presented in Figs.~\ref{f2} and
\ref{f3}.
Fig.~\ref{f2} shows that for each value of $%
S_0$ there are two values of $\Pi_L$ and that the width of the
entropy curve grows inversely with $\theta$. In Fig.~\ref{f3} the
level curves of the entropy as a function of $\theta$ and $\Pi_L$
are presented as projections of the three-dimensional surface. From
these figures it is clear that the maximum of the entropy of the
entanglement is achieved for the classical Markovian process
($\Pi_L=1/2$)

To conclude this section, it is interesting to compare the entropy
of entanglement with the usual Shannon entropy, used in the theory
of communication. In particular one could wonder if the entropy of
the entanglement may be used as a measurement of the degree of
disorder of chirality. The Shannon entropy, in the asymptotic GCD
model, is
\begin{equation}
S_S\equiv-\Pi_{L}\log _{2}\Pi_{L}-\Pi_{R}\log _{2}\Pi_{R},  \label{ss}
\end{equation}
where $\Pi_{L}$ and $\Pi_{R}$ are given by Eq. (\ref{estacio}).

It is clear from Eqs. (\ref{s0},\,\ref{ss}) that when $Q_0=0$
($\Pi_{L}=\Pi_{R}=1/2$) both entropies attain the maximum value
$S_0=S_S=1$. However for other values of $Q_0$ they are different,
in particular, when there is perfect statistical order, \emph{i.e.}
$\Pi_{L}=1$ and $\Pi_{R}=0$ (or $\Pi_{L}=0$ and $\Pi_{R}=1$), the
Shannon entropy vanishes as it should but the entropy of
entanglement does not vanish. Therefore, although the behavior of
the entropy of entanglement is correlated with the behavior of the
GCD, it does not describe correctly the degree of disorder of the
GCD.

\section{Conclusion}

This work provides a different insight for the QW dynamics. It
studies the QW focusing on the probability distribution of the
chirality independently of the position (GCD) and connects this
distribution with the entropy of entanglement. Using an alternative
analytical approach for the QW on the line, developed in previous
works \cite{Alejo2, Alejo3}, we show analytically that the GCD
converges to a stationary solution. The asymptotic behavior of the
GCD looks like the behavior of the two dimensional classical random
walk but, unlike the latter, the asymptotic GCD depends on the
initial conditions. The coexistence of the unitary evolution of the
amplitude together with the asymptotic value of the GCD is a
striking result about the behavior of the system.

We study the entanglement between the coin and the position in the
QW on the line and we show that the behavior of the entropy of
entanglement depends on the GCD. We also show that the asymptotic
entanglement is maximized when the evolution of the GCD follows a
Markovian process. However the entropy of entanglement does not
describe correctly the degree of disorder of the GCD; this is well
described by the Shannon entropy. In previous works
\cite{Carneiro,abal,Maloyer,Annabestani} the dependence of the
asymptotic entropy of entanglement with the initial conditions was
studied, here we provide an analytical recipe to obtain a
predetermined entanglement using extended Gaussian initial
conditions. In other words, starting from a given value of the
entropy of entanglement it is possible to choose the corresponding
initial conditions. These exact expressions can be also used to
obtain a predetermined asymptotic GCD, that is, starting from a
given asymptotic limit of the GCD obtain the corresponding initial
conditions for the QW.

I acknowledge stimulating discussions with V\'{\i}ctor Micenmacher, Guzm\'{a}%
n Hern\'{a}ndez, Ra\'{u}l Donangelo, Eugenio Rold\'{a}n, Germ\'{a}n
J. de Valc\'{a}rcel, Armando P\'{e}rez and Carlos Navarrete Benlloch
and the support from PEDECIBA and ANII.

\end{document}